\documentclass[a4paper]{jpconf}
\usepackage{graphicx}

\newcommand{\zaa}{\emph{Astron.~Astrophys.}}

\newcommand{\zapj}{\emph{Astrophys.~J.}}
\newcommand{\zapjl}{\emph{Astrophys.~J.~Lett.}}
\newcommand{\zapjs}{\emph{Astrophys.~J.~Supp.}}

\newcommand{\znp}{\emph{Nucl.~Phys.}}

\newcommand{\zpr}{\emph{Phys.~Rev.}}
\newcommand{\zprl}{\emph{Phys.~Rev.~Lett.}}

\newcommand{\zmnras}{\emph{Mon. Not. R. Astron. Soc.}}

\newcommand{\zjcap}{\emph{JCAP}}

\newcommand{\ob}{$\Omega_{\mathrm{b}}$}
\newcommand{\obh}{$\Omega_{\mathrm{b}}{\cdot}h^2$}

\newcommand{\deu}{D}
\newcommand{\tro}{$^3$He}
\newcommand{\qua}{$^4$He}
\newcommand{\six}{$^{6}$Li}
\newcommand{\neu}{$^{9}$Be}
\newcommand{\dix}{$^{10}$B}
\newcommand{\onz}{$^{11}$B}

\newcommand{\sep}{$^{7}$Li}

\newcommand{\hli}{$^4$He, D, $^3$He and $^{7}$Li}

\newcommand{\gap}{\mathrel{ \rlap{\raise.5ex\hbox{$>$}}
                    {\lower.5ex\hbox{$\sim$}}  } }
\newcommand{\lap}{\mathrel{ \rlap{\raise.5ex\hbox{$<$}}
	            {\lower.5ex\hbox{$\sim$}}  } }

\newcommand{\sbbn}{Standard Big Bang Nucleosynthesis}
\newcommand{\bbn}{Big Bang  Nucleosynthesis}

\begin{document}
\title{Primordial Nucleosynthesis}

\author{Alain Coc}

\address{Centre de Spectrom\'etrie Nucl\'eaire et de Spectrom\'etrie de
Masse (CSNSM), CNRS/IN2P3, Universit\'e Paris Sud, UMR~8609,
B\^atiment 104, F--91405 Orsay Campus, France}

\ead{Alain.Coc@csnsm.in2p3.fr}

\begin{abstract}
Primordial nucleosynthesis, or \bbn\ (BBN), is one of the three evidences 
for the Big-Bang model, together with the expansion of the Universe
and the Cosmic Microwave Background. There is a good global agreement over a 
range of nine orders of magnitude between
abundances of \hli\  deduced from observations, and calculated in primordial
nucleosynthesis. This comparison was used to determine
the baryonic density of the Universe. For this purpose, it is now 
superseded by the analysis of the Cosmic Microwave Background (CMB) radiation
anisotropies. 
However, there remain, a yet unexplained,  discrepancy of a factor 3--5, between  the calculated 
and observed lithium primordial abundances,
that has not been reduced, neither by recent nuclear physics experiments, nor by new observations.   
We review here the nuclear physics aspects of BBN for the production of \hli, but also \six, \neu,
\onz\ and up to CNO isotopes. These are, for instance, important for the initial composition of the 
matter at the origin of the first stars.
Big-Bang nucleosynthesis, that has been used, to first constrain the baryonic density, and the number of
neutrino families, remains, a valuable tool to probe the physics of the early Universe, like
variation of "constants" or alternative theories of gravity.

\end{abstract}

\section{Introduction}
\label{s:intro}
\bigskip

There are presently three evidences for the Big-Bang Model : the universal expansion, the Cosmic Microwave Background (CMB) radiation
and Primordial or Big-Bang Nucleosynthesis (BBN).
The third evidence for a hot Big-Bang comes, indeed, from the primordial
abundances of the ``light elements'': \hli. They were produced during the 
first $\approx$20 minutes of the Universe when it was dense and hot enough for
nuclear reactions to take place. 
These primordial abundances are compared to astronomical observations 
in primitive astrophysical sites.
It is worth reminding that \bbn\ has been essential in the past, to first estimate the baryonic density 
of the Universe, $\rho_{\rm B} = (1-3)\times10^{-31}$~g/cm$^3$ \cite{Wag73}, and give an upper limit 
on the  number neutrino families $N_\nu\leq3$\cite{Yan79}, both in the seventies.
The number of light neutrino families is now known from the measurement of the $Z^0$ 
width by LEP experiments at CERN: $N_\nu$ = 2.9840$\pm$0.0082~\cite{LEP}. 
The nuclear reaction rates have all been measured in nuclear physics laboratories or can be calculated 
from the standard theory of weak interactions. In that case, they are normalized to the experimental value for the 
lifetime of the neutron. Its precise value is still a matter of debate \cite{Wie11} 
$\tau_{\mathrm{n}}$ = 880-884~s, but its uncertainty has only marginal effect on BBN.
The last parameter to have been independently determined is the precise value of baryonic density 
of the Universe, which is now deduced from the observations of the anisotropies 
of the CMB radiation.
It is usual to introduce $\eta$, the number of photons per baryon which
remains constant during the expansion, and is directly related to \ob\ by
\obh=3.65$\times10^7\eta$ with 
\begin{equation}
\Omega_{\mathrm{b}}{\cdot}h^2=0.02249\pm0.00062  \; {\rm and} \; \Omega_{\mathrm{b}}=0.04455\pm0.0027  
\label{eq:omega}
\end{equation} 
 (``WMAP only Seven Year Mean'' \cite{WMAP}). The parameter $h$ 
represents the Hubble constant ($H_0$) in units of 100~km/s/Mpc (i.e. $h$= 0.704  \cite{WMAP}) and \ob$\equiv\rho_{\rm B}/\rho_{\rm 0,C}$ the baryonic density relative to 
the {\em critical density}, $\rho_{\rm 0,C}$, which corresponds to a flat (i.e. Euclidean) space. It is given by :
\begin{equation}
\rho_{\rm 0,C}={{3H_0^2}\over{8{\pi}G}}=1.88\;h^2\times10^{-29}\;
\mathrm{g/cm^3}\;\mathrm{or}\;2.9\;h^2\times10^{11}\;
\mathrm{M_\odot}/\mathrm{Mpc^3}
\label{eq:critic}
\end{equation} 
where $G$ is the gravitational constant. It corresponds to a density of
a few hydrogen atoms per cubic meter or one typical galaxy per cubic
megaparsec (Mpc).
This results (Eqs.~\ref{eq:omega}--\ref{eq:critic}) in a baryonic density which is just slightly above the range provided by Wagoner \cite{Wag73} in 1973!

Hence, the number of free parameters in \sbbn\ has now been reduced to  zero, and the calculated 
primordial abundances are in principle only affected by the moderate uncertainties in
some nuclear cross--sections. It may appears that \bbn\ studies are now useless, but this is certainly
not the case. First, even though the agreement with observations is good or very good for \qua, \tro\ and
\deu, there is a tantalizing discrepancy for \sep\ that has not yet found a consensual explanation.
Second, when we look back in time, it is the ultimate process for which, {\it a priori}, we know all the 
physics involved. Hence,  departure from its predictions could provide hints 
or constraints on new physics or astrophysics.

Besides the \hli\ isotopes, some minute traces of \six, \neu, \onz\ and CNO are produced by BBN.
Observations of \six\ in a few halo stars have renewed the interest for this isotope and the nuclear
uncertainties concerning its production. The evolution of the first generation (Population III) of stars
could be influenced by the amount of primordial CNO elements, as hydrogen burning can proceed
either through the slow p--p chain, or through the more efficient CNO cycle. We will hence review the nuclear
aspects of the primordial production of element up to oxygen.

\section{The cosmological elements \hli}
\bigskip

\subsection{Abundances of the cosmological elements}
\label{s:abund}
\bigskip

During the evolution of the Galaxy, complex nucleosynthesis takes place, mainly
in massive stars which release matter enriched in heavy elements into the 
interstellar medium when they explode as supernovae. 
Accordingly, the abundance of heavy elements in the gas, at the origin of star
formation, increases with time. 
The observed abundance of 
{\em metals} (in astrophysics, the chemical elements beyond helium)
is hence an indication of its age: the oldest stars have the lowest metallicity.
To derive the most primitive abundances one has first, to extract them from observations of astrophysical sites  which
are thought to be non evolved and second, extrapolate them  to zero metallicity.

Primordial lithium abundance is deduced from observations of low metallicity stars in the halo of our Galaxy
where the lithium abundance is almost independent of metallicity, displaying a plateau \cite{Spi82}. 
This constant Li abundance is interpreted as corresponding to the BBN \sep\  yield.
This interpretation assumes that lithium has not been depleted at the 
surface of these stars so that the presently observed abundance is supposed to be equal 
to the initial one. The small scatter of values around the ``Spite plateau'' is an indication 
that depletion may not have been very effective.
Astronomical observations of metal poor halo stars have led to a relative primordial abundance \cite{Sbo10} of:
\begin{equation}   
\hbox{Li/H} =  (1.58 \pm 0.31) \times 10^{-10}.
\end{equation}

Note also that observationally challenging detections of \six\ has been reported \cite{Asp06} to a level of $\sim10^{-2}$
below the  Spite plateau value. The presence of a \six\ plateau is however not confirmed as only few (2-3) stars seem
to present significant \six\ abundances on their surfaces \cite{Ste12} .
For a recent review of the latest Li observations and their different astrophysical aspects, see Refs.~\cite{Spi10,Ioc12}, and the proceedings
of the  2012 Workshop {\em Lithium in the Cosmos}\footnote{{\tt http://www.iap.fr/lithiuminthecosmos2012/index.html}, proceedings will appear in
{\it Memorie della Societ\`a Astronomica Italiana Supplementi}.}.

Contrary to \sep\ which can be both produced (spallation,
asymptotic giant branch (AGB) stars, novae)
and destroyed (in the interior of stars), deuterium is a very fragile
isotope, that can only be destroyed after BBN. 
Its most primitive abundance is determined from the observation of absorption lines in
clouds at high redshift, on the line of sight of more
distant quasars.  
Very few observations of these
cosmological clouds  are available and a weighted mean \cite{Oli12} (and references therein) of this data yields a D/H abundance of: 
\begin{equation}
\hbox{D/H} = (3.02 \pm 0.23) \times 10^{-5}. 
\end{equation}

After BBN, \qua\ is produced by stars. Its primitive abundance is 
deduced from observations in HII (ionized hydrogen) regions of compact 
blue galaxies. 
Galaxies are thought to be formed by the agglomeration of such dwarf galaxies
which are hence considered as more primitive.
The primordial \qua\ abundance $Y_{\mathrm{p}}$ (\qua\ mass fraction) is
given by the extrapolation to zero metallicity  but is affected by systematic 
uncertainties such as plasma 
temperature or stellar absorption.
Using the data compiled in Ref.~\cite{Izo07}, it was found \cite{Ave11,Ave12} that:
\begin{equation}
\hbox{Y}_p = 0.2534 \pm 0.0083.
\label{eq:qua}
\end{equation}

Contrary to \qua, \tro\ is both produced and destroyed in stars so that
the evolution of its abundance as a function of time is not well known, and  has only been observed in our Galaxy \cite{Ban02}: 
\begin{equation}
\hbox{\tro/H}= (1.1 \pm 0.2) \times 10^{-5}.
\end{equation}
Consequently, comparison with \tro\ abundance from BBN is subject to caution \cite{Van03}.

\subsection{Nuclear reactions for \hli\ nucleosynthesis}
\label{s:nucl}
\bigskip

Unlike in other sectors of nuclear astrophysics, nuclear cross 
sections have usually been directly measured at BBN energies
($\sim$100~keV).  There are 12 nuclear reactions
responsible for the production of ~\hli\ in Standard BBN. 
There are many other reactions connecting these isotopes, but their
cross sections are too small and/or reactants too scarce to have 
any significant effect.  Even among these 12 reactions, a few of them
are now irrelevant (see below) at WMAP baryonic density.

The weak reactions involved in n$\leftrightarrow$p equilibrium are
an exception; their rates \cite{Dic82} come from the standard theory
of the weak interaction, normalized to the experimental neutron 
lifetime. 
Until very recently, the averaged value of 885.7$\pm$0.8~s was recommended by the Particle Data Group, but
new measurement lead to significantly lower values. 
While it has not yet been possible to solve this discrepancy \cite{Wie11}, 
a reevaluation of the recommended value:  880.1$\pm$1.1~s has been proposed \cite{PDG}, 
awaiting experimental clarification.

The $^1$H(n,$\gamma)^2$H cross section is also obtained from
theory \cite{And06} but in the framework of Effective Field Theory.


\emergencystretch30pt
For the ten remaining reactions, $^2$H(p,$\gamma)^3$He, $^2$H(d,n)$^3$He, 
$^2$H(d,p)$^3$H, $^3$H(d,n)$^4$He, 
$^3$H($\alpha,\gamma)^7$Li, $^3$He(d,p)$^4$He, $^3$He(n,p)$^3$H, 
$^3$He($\alpha,\gamma)^7$Be, $^7$Li(p,$\alpha)^4$He and $^7$Be(n,p)$^7$Li,
the cross sections have been measured in the laboratory at the relevant energies.
Formerly, we used the reaction rates from the the evaluation 
performed by Descouvemont et al. \cite{Des04}.
However, more recent experiments and analysis have lead to improved reaction rates for several 
important reactions.

\begin{table}[tb]
\caption{\label{t:sensitivity}Abundance sensitivity to reaction rates: ${\partial\log}Y/{\partial\log}<{\sigma}v>$ at WMAP baryonic density.} 
\begin{center}
\lineup
\begin{tabular}{cccccc}
\br                              
Reaction &  $^4$He & D & $^3$He & $^7$Li & $E_0$( ${\Delta}E_0/2$) \cr 
                 &                  &     &                &               &     (MeV) \cr
\mr
n$\leftrightarrow$p ($\tau_n^{-1}$)& -0.73 & 0.42 & 0.15 & 0.40 & \cr 	
$^1$H(n,$\gamma)^2$H & 0 & -0.20 & 0.08 & 1.33	 & \cr 
$^2$H(p,$\gamma)^3$He & 0	& -0.32 & 0.37 & 0.57 & 0.11(0.11) \cr
$^2$H(d,n)$^3$He & 0 & -0.54 & 0.21 & 0.69 & 0.12(0.12) \cr
$^2$H(d,p)$^3$H    & 0 & -0.46 & -0.26 & 0.05 & 0.12(0.12) \cr
$^3$H(d,n)$^4$He & 0 & 0 & -0.01 & -0.02 & 0.13(0.12) \cr
$^3$H($\alpha,\gamma)^7$Li & 0 & 0 & 0 & 0.03 & 0.23(0.17) \cr
$^3$He(n,p)$^3$H & 0 & 0.02 & -0.17 & -0.27 & \cr	
$^3$He(d,p)$^4$He & 0 & 0.01 & -0.75 & -0.75 & 0.21(0.15) \cr
$^3$He($\alpha,\gamma)^7$Be &  0 & 0 & 0 & 0.97 & 0.37(0.21) \cr
$^7$Li(p,$\alpha)^4$He & 0 & 0 & 0 & -0.05 & 0.24(0.17) \cr
$^7$Be(n,p)$^7$Li & 0 & 0 & 0& -0.71 & \cr	
\br
\end{tabular}
\end{center}
\label{t:sensib}
\end{table}

To point out the most important reactions, we display in Table~\ref{t:sensib} 
the sensitivity  of the calculated abundances ($Y_i$ with $i$ = \hli)  w.r.t. to a change in the 12 
reaction rates by a constant factor. We define the sensitivity as ${\partial\log}Y/{\partial\log}<{\sigma}v>$,
based on the assumption that the nuclear cross section uncertainties are now dominated
by systematic uncertainties that affect their normalization rather than by statistics.
These values were obtained, at the WMAP baryonic density, by a parabolic fit of ${\Delta}Y_i/Y_i$ for $\pm$15\% 
variations of the reaction rates. 
The last column represents the Gamow window at BBN  typical temperatures.

This table can be used as a guide for further experimental efforts. We see for instance that {\em at WMAP baryonic
density}, the  $^3$H($\alpha,\gamma)^7$Li and $^7$Li(p,$\alpha)^4$He reactions
play a negligible role. At WMAP baryonic density, \sep\ is produced indirectly by $^3$He($\alpha,\gamma)^7$Be,
that will, much later decay to \sep.

The sensitivity to the weak rates is high but (within standard theory), the uncertainty 
is governed by the neutron lifetime which is now known with reasonable precision. 
Indeed, the relative uncertainty on $\tau_n$  is of the order of 5$\times10^{-3}$ \cite{Wie11}
affects \qua\ abundance by (Table~\ref{t:sensib}) $-0.73\times5\times10^{-3}$, a factor of ten lower than the
observational uncertainty (Eq.~\ref{eq:qua}).

The influence of the  $^1$H(n,$\gamma)^2$H rate was unexpected. The
\sep\ final abundance depends strongly on the rate of this reaction while other isotopes are little affected. 
This unexpected effect can be traced to the increased neutron abundance at
$^7$Be formation time for a low   $^1$H(n,$\gamma)^2$H rate making its destruction by
neutron capture, $^7$Be(n,p)$^7$Li(p,$\alpha)^4$He, more efficient (see Fig.~1 in \cite{Coc07}).
However,  the few experimental informations available for this cross section at BBN energies
are in good agreement with the calculations estimated to be reliable to within 1\% error\cite{And06}.

\begin{figure}
\begin{center}
  \includegraphics[width=\textwidth]{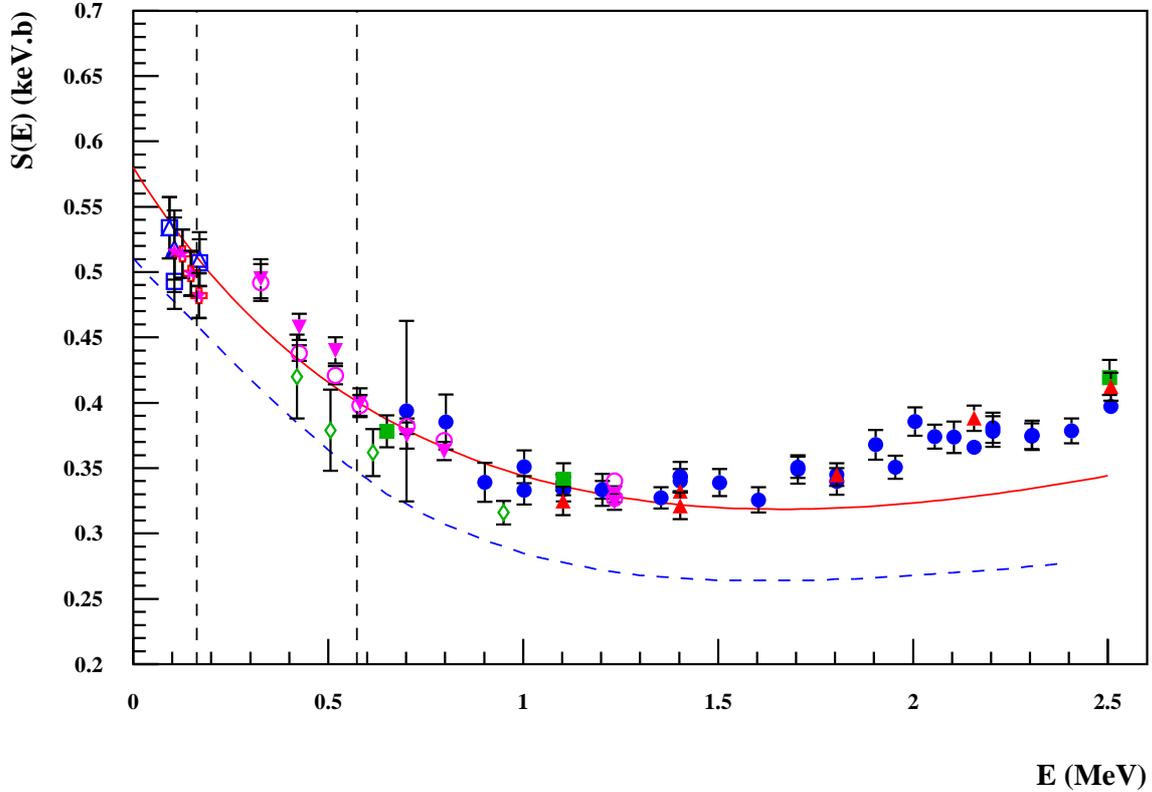}
  \caption{Astrophysical S--factor for the $^3$He($\alpha,\gamma)^7$Be reaction, adapted from Fig.~3 in Di Leva et al. \cite{DiL09}. 
  The data are from Refs.~\cite{hag,DiL09},  the 
  dot--dashed and solid curves are respectively the previously used Descouvemont et al. \cite{Des04} (anterior to the displayed data \cite{hag,DiL09})
  or the adopted Cyburt \& Davids
  \cite{Cyb08a} (including recent data \cite{hag}) fits. The dashed vertical lines represent the Gamow window at 1~GK.}
\label{f:hag}
\end{center}
\end{figure}

The next most important reaction (Table~\ref{t:sensib}) is  $^3$He($\alpha,\gamma)^7$Be as it is the 
path for the formation of \sep\ at high density. Hence, the \sep\ abundance is directly proportional 
to this rate, which has long been a subject of debate. Systematic differences in the measured cross 
section were found according to the experimental technique: prompt or activation measurements.
Thanks to the recent experimental efforts \cite{hag,DiL09}, 
in particular at  LUNA at the Laboratori Nazionali del Gran Sasso,
the two methods provide now results in agreement, within each others error bars.
With this new experimental data, Cyburt \& Davids~\cite{Cyb08a} calculated the S--factor 
which is  significantly higher than the Descouvemont et al. \cite{Des04} R--matrix fit,
done before these new data were available (Fig.~\ref{f:hag}). 
This explains the higher \sep\ primordial abundance obtained in recent calculations.
At high energy, the recent experimental data,
in particular of Di Leva et al. \cite{DiL09}, obtained by a third technique, the recoil mass separation, deviate from both fits.
Theoretical explanations are available \cite{Nef11}, but this should not affect the S--factor at BBN energies.  
Nevertheless, one may note the relative scarcity of experimental data within the Gamow window (Fig.~\ref{f:hag}).

\begin{figure}[htb]
\begin{center}
\includegraphics[width=\textwidth]{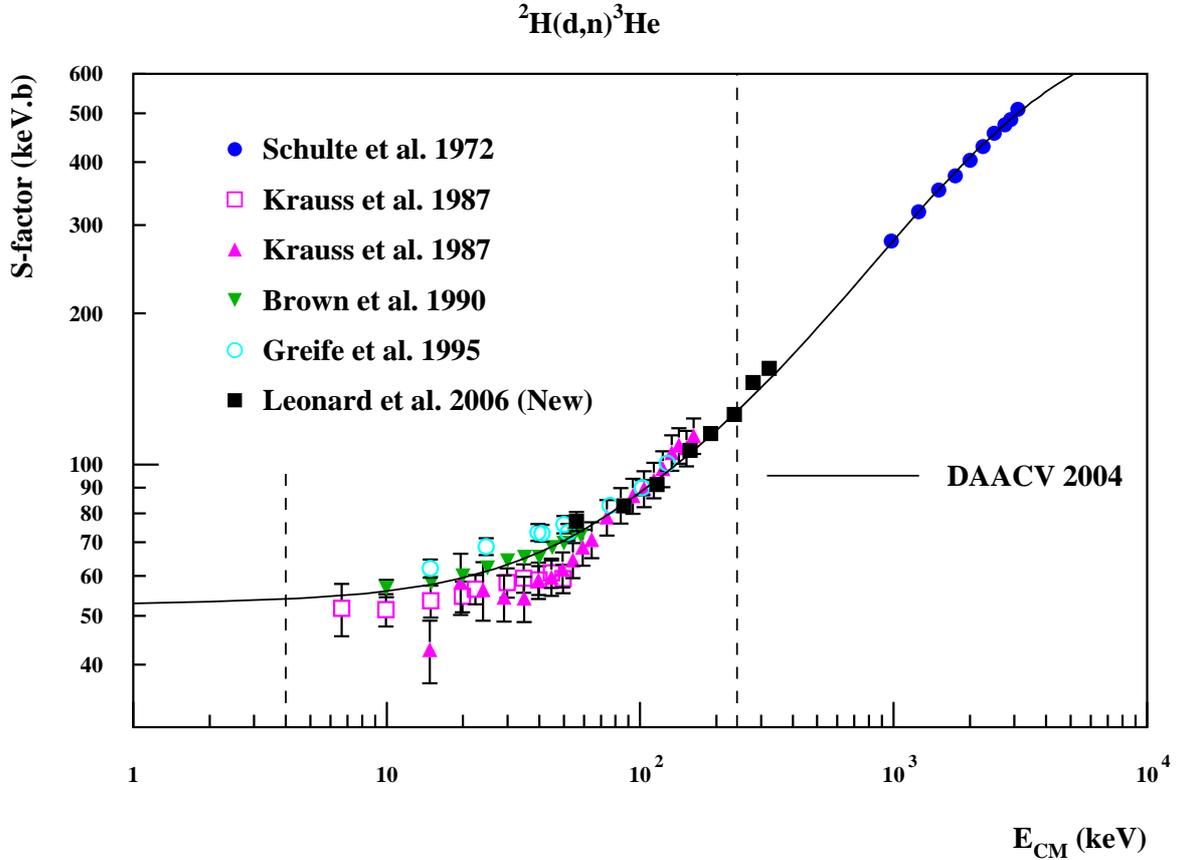}
\caption{Experimental data for the, $^2$H(d,n)$^3$He reaction, $S$--factor compared with
the R--matrix fit of Descouvemont et al. \cite{Des04}. (See this reference for reference to 
experimental data except for the more recent data from Leonard et al. \cite{Leo06}.) The dashed vertical lines represent the Gamow window at 1~GK} 
\label{f:ddn}
\end{center}
\end{figure}

The $^2$H(d,n)$^3$He reaction, also influential on \sep, was re-measured [together with $^2$H(d,p)$^3$H]
by Leonard \etal\cite{Leo06}, after the  R-matrix analysis \cite{Des04} was performed. The very precisely measured 
cross section is in perfect agreement with the R-matrix fit (Fig.~\ref{f:ddn}).

\subsection{BBN primordial abundances compared to observations}
\bigskip

\begin{figure}
\begin{center}
\vskip -2cm
  \includegraphics[height=.9\textheight]{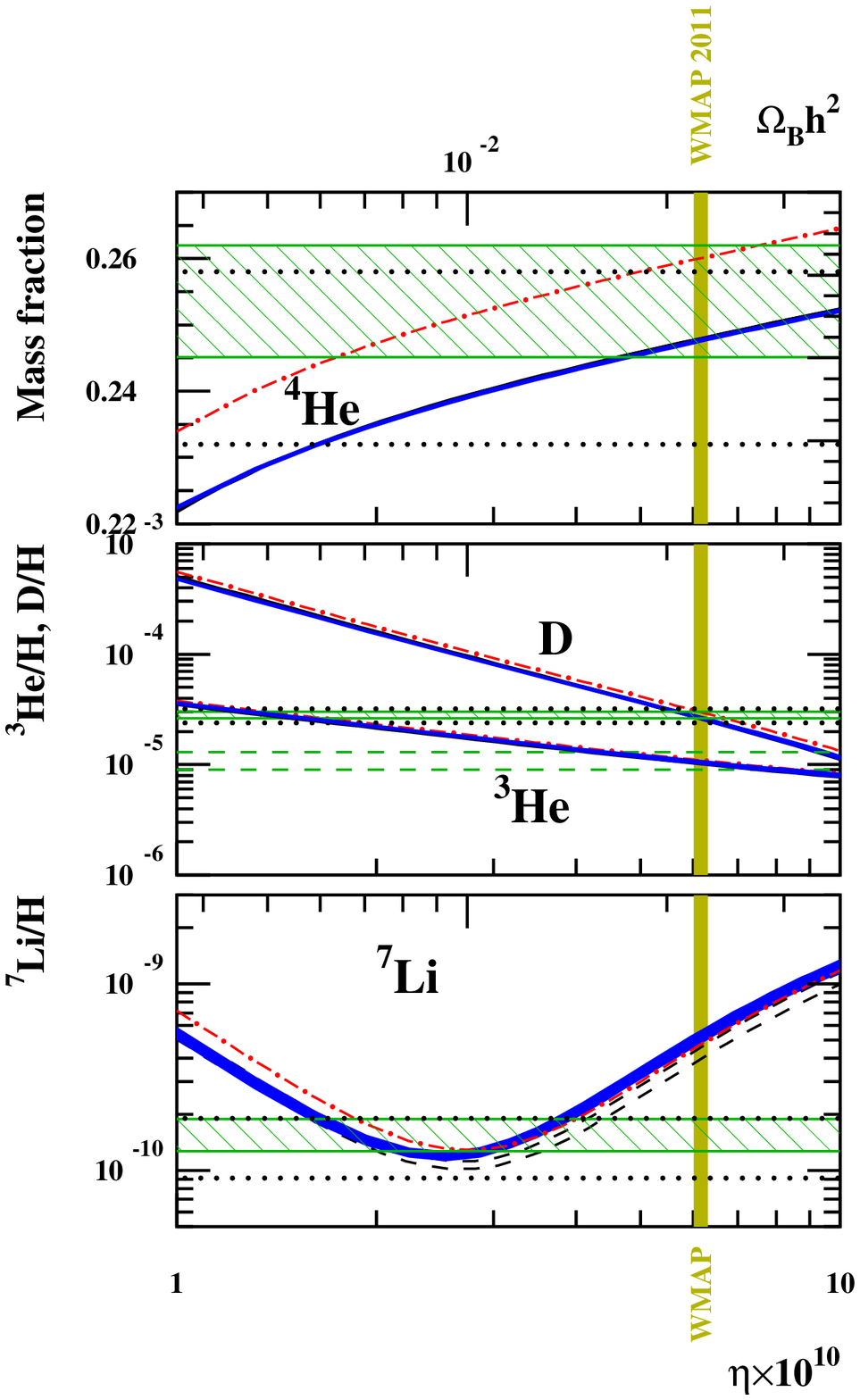}
\caption{\label{f:heli}
Abundances of \qua\ (mass fraction), \deu, \tro\ and \sep\ (by
number relative to H) as a function of the baryon over photon ratio 
$\eta$ (or \obh.) showing the effect of nuclear uncertainties \cite{CV10}. 
The vertical stripe corresponds to the WMAP baryonic density \cite{WMAP}
while the horizontal area represent the adopted primordial abundances (dotted lines 
those adopted in CV10~\cite{CV10}).
The dashed curves 
represent previous calculations \cite{Coc04} before the re-evaluation \cite{Cyb08a} 
of the $^3$He($\alpha,\gamma)^7$Be rate. The  dot-dashed lines corresponds, 
to  4 {\em effective} neutrino families.}
\end{center}
\end{figure}

Figure~\ref{f:heli} shows the abundances of 
\qua\ (mass
fraction), \deu, \tro\ and \sep\ (in number of atoms relative to H)
as a function of the baryonic density. 
The thickness of the curves reflect the nuclear uncertainties. 
They were obtained by a Monte-Carlo calculation using
for the nuclear rate uncertainties those obtained by \cite{Des04}
with the notable exception of $^3$He($\alpha,\gamma)^7$Be \cite{Cyb08a} and 
$^1$H(n,$\gamma)^2$H \cite{And06}.
The horizontal lines represent the limits on the \qua, \deu\ and \sep\
primordial abundances deduced from spectroscopic observations.
The vertical stripe represents the baryonic density deduced from
CMB observations  \cite{WMAP}.
The concordance between BBN and observations is
 in perfect agreement for deuterium. Considering the large 
uncertainty associated with \qua\ observations, the agreement with
CMB+BBN is fair. The calculated \tro\ value is close to its 
galactic value showing that its abundance has little changed
during galactic chemical evolution.    
On the contrary, the \sep, CMB+BBN calculated abundance is significantly
higher than the spectroscopic observations : from a factor of $\approx$3\cite{Coc04} when 
using the Descouvemont et al. library \cite{Des04} only and the Ryan 
et al. observations \cite{Rya00} (dotted lines in lower panel of Fig.~\ref{f:heli}), to a factor of $\approx$5 \cite{Cyb08b,CV10} when
using the new rates and Li observations \cite{Sbo10}.
Table~\ref{t:yields} displays the comparison between BBN
 abundances deduced from the WMAP results and
  the spectroscopic observations.
The origin of this lithium discrepancy between
CMB+BBN and spectroscopic observations remains an open question.
Note also that with the new determination of the \qua\ primordial abundances \cite{Ave11,Ave12}, the agreement 
for \qua\ becomes marginal. As shown on the figure, an increase of the rate of expansion of the universe during BBN
(simulated by an additional effective neutrino family) would improve the situation.  

\begin{table}[h]
\caption{\label{t:yields} Yields at WMAP baryonic density compared to observations.} 
\begin{center}
\lineup
\begin{tabular}{ccccc}
\br                              
& Cyburt el al 2008\cite{Cyb08b} & CV10 \cite{CV10}    & Observations & Factor\cr 
\mr
$^4$He     & 0.2486$\pm$0.0002   & 0.2476$\pm$0.0004 & 0.2534$\pm$0.0083\cite{Ave12}    & $\times10^0$ \cr
D/H            & 2.49$\pm$0.17	            & 2.68$\pm$0.15          & $3.02 \pm 0.23$ \cite{Oli12}  & $\times10^{-5}$ \cr
$^3$He/H & 1.00$\pm$0.07	            & 1.05$\pm$0.04          & 1.1$\pm$0.2\cite{Ban02}          & $\times10^{-5}$ \cr
$^7$Li/H   & 5.24$^{+0.71}_{-0.67}$ &5.14$\pm$0.50           & 1.58$\pm$0.31\cite{Sbo10}    & $\times10^{-10}$ \cr
\br
\end{tabular}
\end{center}
\end{table}

\section{BBN production of heavier elements}
\bigskip

Even though the direct detection of primordial CNO isotopes
seems highly unlikely with the present observational techniques at high redshift, 
it is important to better estimate their \sbbn\ production.
Hydrogen burning in the first generation of stars (Pop III stars) proceeds through the
slow p--p chains until enough carbon is produced (through the triple-alpha reaction)
to activate the CNO cycle. The minimum value of the initial CNO abundance that
would affect Pop III stellar evolution is estimated to be as low as  10$^{-13}$ 
(in number of atoms relative to hydrogen, CNO/H) for the less massive ones \cite{Eks08}.
This is only two orders of magnitude above the  \sbbn\ CNO yield, using the current
nuclear reaction rate evaluations of Iocco et al. \cite{Ioc07}.
In addition, it has been shown that Pop III stars evolution is sensitive to the triple-alpha
($^{12}$C producing) reaction and can be used to constrain the possible variation of the
fundamental constants \cite{Eks10}. This reaction rate is very sensitive to the position of the Hoyle
state, which in turn is sensitive to the values of the fundamental constants. 
The same mechanism could also increase the amount of CNO ($^{12}$C) produced in BBN.
In the same context of the
variations of the fundamental constants, $^8$Be (which decays to two alpha particles
within $\sim10^{-16}$~s) could become stable if these constants were only slightly
different. At BBN time, this would possibly allow to bridge the "A=8 gap" and produce
excess CNO \cite{Coc12b}. To determine how significant would be this excess, one needs to know
the standard  BBN production of the CNO elements.

The production of  CNO isotopes has been studied in the context of standard and
inhomogeneous BBN.   
The most relevant analysis comes from  Iocco et al. \cite{Ioc07} who included more than 100 nuclear reactions and predicted 
a CNO/H abundance ratio  of approximately $6 \times 10^{-16}$, with an upper limit of  $10^{-10}$.

The main difficulty in BBN calculations up to CNO is the extensive network needed,
including n-, p-, $\alpha$-, but also d-, t- and  \tro-induced reactions. Most of the corresponding
cross sections cannot be extracted from experimental data only. This is especially
true for radioactive tritium-induced reactions, or for those involving radioactive targets like
e.g. $^{10}$Be. For some reactions, experimental data, including spectroscopic
data of the compound nuclei, are just nonexistent. Hence, for many reactions,
one has to rely on theory to estimate the reaction rates. Previous studies lack documentation
on the origin of the reaction rates, but
have apparently extensively used old and unreliable prescriptions to estimate many of them. 
A detailed analysis of all reaction rates and associated uncertainties would be
desirable but is impractical for a network of $\approx$400 reactions.
So we first performed a sensitivity study to identify the 
most important reactions, followed by dedicated re-evaluations. 
For this purpose, we used, as a first guess, the more reliable reaction rate estimates provided 
by the TALYS code \cite{TALYS}.

\subsection{Sensitivity study}
\bigskip

To extend our BBN network, we need the neutron, proton, deuterium, tritium,$^3$He and $\alpha$-particle capture cross sections 
on targets in the A=1 to 20 range. 
As a first approximation, we used the results from TALYS for 
rates that are not available in the literature
(the full list of references can be found in Ref.~\cite{Coc12a}).
By comparing TALYS results with experimentally determined reaction rates \cite{NACRE,Ili10b}, we observed that, even for very light elements like Li, 
TALYS  globally  provides predictions, "accurate" within 3 orders of magnitude, in the temperature range of interest here.
Hence variations of these theoretical rates by up to three orders of magnitude can in a first step be used in our sensitivity analysis.
Sensitivity studies have already been performed for the reactions involved in \hli\ production \cite{Cyb04,Ser04,Coc04} but 
here, we will concentrate on the C, N and O isotopes production.
To estimate the impact of the reaction rate uncertainties on \sbbn,
we perform for each reaction six additional calculations,  changing its rate by factors of 0.001, 0.01, 0.1, 10, 100
and 1000, and calculate the relative change in CNO abundances.  (Mass fractions of isotopes with A$\geq$12 are added together into CNO.)
Tables~\ref{t:sensibcno} to \ref{t:sensibonz}  display, reactions for which the relative changes in \six, \neu, \onz\ and CNO abundances 
are larger than 20\%.
The last column contains the reference for the origin of the 
reaction rate used {\em for the sensitivity study} (the TALYS rates are available electronically \cite{web}); i.e.
before reaction rate re-evaluation (next section).

\begin{table}[h]
\begin{center}
\begin{tabular}{ccccccccc}
\hline
 Factors & 0.001 & 0.01 & 0.1 & 10. & 100. & 1000. & $\times<{\sigma}v>$ & \\
\hline
Reaction &\multicolumn{6}{c}{Fractional change in CNO} & Ref.        \\
\hline
$^7$Li(d,$\gamma)^9$Be & 1.00 & 1.00 & 1.00 & 1.01 & 1.11 & 2.10 & TALYS \\
$^7$Li(d,n)2$\alpha$ & 1.66 & 1.65 & 1.55 & 0.28 & 0.06 & 0.02 & \cite{Boy93} \\
$^7$Li(t,n)$^9$Be & 0.99 & 0.99 & 0.99 & 1.10 & 2.14 & 11.7 & \cite{Bru91} \\
$^7$Li(t,2n)2$\alpha$ & 1.00 & 1.00 & 1.00 & 0.99 & 0.91 & 0.53 & \cite{MF89} \\
$^8$Li(n,$\gamma)^9$Li & 1.00 & 1.00 & 1.00 & 1.01 & 1.06 & 1.62 & \cite{Rau94} \\
$^8$Li(t,n)$^{10}$Be & 1.00 &  1.00 & 1.00 & 1.00 & 1.02 & 1.23 & TALYS \\
$^8$Li($\alpha,\gamma)^{12}$B  & 1.00 & 1.00 & 1.00 & 1.01 & 1.11 & 2.15 & TALYS \\
$^8$Li($\alpha$,n)$^{11}$B & 0.89 & 0.89 &0.90 & 1.97 & 11.2 & 78.1 & \cite{Miz00} \\
$^9$Li($\alpha$,n)$^{12}$B &  1.00 & 1.00 & 1.00 & 1.01 & 1.08 & 1.73 & TALYS \\
$^{10}$Be($\alpha$,n)$^{13}$C & 1.00 & 1.00 & 1.00 & 1.00 & 1.03 & 1.28 & TALYS \\
$^{11}$B(n,$\gamma)^{12}$B & 0.91 & 0.91& 0.92 & 1.81 & 9.91 & 87.7 & \cite{Rau94} \\
$^{11}$B(d,n)$^{12}$C & 0.70 & 0.71 & 0.73 & 3.67 & 30.2 & 280. & TALYS \\
$^{11}$B(d,p)$^{12}$B & 0.99 & 0.99 & 0.99 & 1.08 & 1.83 & 9.33 & TALYS \\
$^{11}$B(t,n)$^{13}$C &1.00 & 1.00 & 1.00 & 1.01 & 1.12 & 2.17 & TALYS \\
$^{11}$C(n,$\gamma)^{12}$C & 1.00 & 1.00 & 1.00 & 1.01 & 1.08 & 1.75 & \cite{Rau94} \\
$^{11}$C(d,p)$^{12}$C & 0.99 & 0.99 & 0.99 & 1.05 & 1.55 & 5.67 & TALYS \\
$^{12}$C(t,$\alpha)^{11}$B & 1.00 & 1.00 & 1.00 & 1.00 & 0.97 & 0.75 & TALYS \\
$^{13}$C(d,$\alpha)^{11}$B & 1.00 & 1.00 & 1.00 & 0.96 & 0.84 & 0.75 & TALYS \\
\hline
\end{tabular}
\caption{\label{t:sensibcno}
Sensitivity of the most important reactions for CNO production in BBN,
to reaction rate variations around initial test rates from references (last column).}
\end{center}
\end{table}

\begin{table}[h]
\begin{center}
\begin{tabular}{ccccccccc}
\hline
 Factors & 0.001 & 0.01 & 0.1 & 10. & 100. & 1000. & $\times<{\sigma}v>$ & \\
\hline
Reaction &\multicolumn{6}{c}{Fractional change in CNO} & Ref.        \\
\hline
$^3$He(t,$\gamma)^6$Li & 1.00 & 1.00 & 1.00 & 1.03 & 1.31 & 4.11 & \cite{FK90} \\
$^4$He(d,$\gamma)^6$Li & 0.004 & 0.013 & 0.010 & 9.97 & 99.7 & 995. & \cite{Ham10} \\
\hline
\end{tabular}
\caption{\label{t:sensibsix}
Same as Table~\ref{t:sensibcno} but for \six.}
\end{center}
\end{table}

\begin{table}[h]
\begin{center}
\begin{tabular}{ccccccccc}
\hline
 Factors & 0.001 & 0.01 & 0.1 & 10. & 100. & 1000. & $\times<{\sigma}v>$ & \\
\hline
Reaction &\multicolumn{6}{c}{Fractional change in CNO} & Ref.        \\
\hline
$^7$Li(d,$\gamma)^9$Be & 0.83 & 0.83 & 0.85 & 2.52 & 17.7 & 170 & TALYS \\
$^7$Li(t,n)$^9$Be & 0.52 & 0.53 & 0.57 & 5.29 & 48.2 & 477. & \cite{Bru91} \\
$^7$Li($^3$He,p)$^9$Be & 1.00 & 1.00 & 1.00 & 1.04 & 1.45 & 5.49 & TALYS \\
$^7$Be(d,p)2$\alpha$ & 1.01 & 1.01 & 1.01 & 0.95 & 0.67 & 0.38 & \cite{CF88} \\
$^7$Be(t,p)$^9$Be  & 0.65 & 0.65 & 0.69 & 4.15 & 35.6 & 345. & TALYS \\
\hline
\end{tabular}
\caption{\label{t:sensibneu}
Same as Table~\ref{t:sensibcno} but for \neu.}
\end{center}
\end{table}

\begin{table}[h]
\begin{center}
\begin{tabular}{ccccccccc}
\hline
 Factors & 0.001 & 0.01 & 0.1 & 10. & 100. & 1000. & $\times<{\sigma}v>$ & \\
\hline
Reaction &\multicolumn{6}{c}{Fractional change in CNO} & Ref.        \\
\hline
$^3$He(t,np)$^4$He & 1.00 & 1.00 & 1.00 & 1.00 & 0.97 & 0.79 & \cite{CF88}  \\
$^7$Be(d,p)2$\alpha$ & 1.01 & 1.01 & 1.01 & 0.93 & 0.55 & 0.11 & \cite{CF88} \\
$^{11}$C(n,$\alpha$)2$\alpha$  & 1.16 & 1.16 & 1.15 & 0.40 & 0.01 & 0.0001 & \cite{Rau94} \\
\hline
\end{tabular}
\caption{\label{t:sensibonz}
Same as Table~\ref{t:sensibcno} but for \onz.}
\end{center}
\end{table}

The examination of Table~\ref{t:sensibcno} shows that, among the $\approx$400,  only a few reactions have a
strong impact on the CNO final abundance. 
The CNO production is significantly sensitive (more than by a factor of about 2) to several reaction rates. 
In particular, these include:
$^7$Li(d,n)2$^4$He, $^7$Li(t,n)$^9$Be,  $^8$Li($\alpha$,n)$^{11}$B,
$^{11}$B(n,$\gamma)^{12}$C, $^{11}$B(d,n)$^{12}$C, 
$^{11}$B(d,p)$^{12}$B and $^{11}$C(d,p)$^{12}$C.
The impact of $^7$Li(d,n)2$^4$He is unexpected and should be compared to the 
influence of $^1$H(n,$\gamma)^2$H on \sep\ (see \cite{Coc07}).
Indeed, when increasing the $^7$Li(d,n)2$^4$He reaction rate by a factor of 1000,
even though the  \hli\  {\em final abundances are left unchanged}, the peak \sep\ abundance
at $t\approx$200~s is reduced by a factor of about 100 (see Fig.~15 in \cite{Coc12a}), an evolution followed by
$^8$Li and CNO isotopes. 
From this table, we can deduce that
the main nuclear paths to CNO (see also \cite{Ioc07}) proceeds from the $^7$Li($\alpha,\gamma)^{11}$B reaction followed
by    $^{11}$B(p,$\gamma$)$^{12}$C, $^{11}$B(d,n)$^{12}$C, $^{11}$B(d,p)$^{12}$B and 
$^{11}$B(n,$\gamma$)$^{12}$B reactions. Another nucleosynthesis path starts with  
$^7$Li(n,$\gamma)^8$Li($\alpha$,n)$^{11}$B. 
(Note that primordial \onz\ is produced by a different path: the late decay of $^{11}$C.)

The examination of Tables~\ref{t:sensibsix} to \ref{t:sensibonz} show that the most important reactions for 
\six, \neu\ and \onz\ productions are $^4$He(d,$\gamma)^6$Li, $^7$Li(t,n)$^9$Be, $^7$Be(t,p)$^9$Be, 
$^7$Li(d,$\gamma)^9$Be and $^{11}$C(n,$\alpha$)2$\alpha$.
However, some of them have been measured ($^4$He(d,$\gamma)^6$Li \cite{Ham10} and $^7$Li(t,n)$^9$Be 
\cite{Bru91,Bar91}) so that the uncertainties on their cross sections are small compared to the range 
explored in this sensitivity study. Considering the tiny production of these isotopes in {\em Standard} BBN
(next section), compared to present day observational techniques, these uncertainties are not important.

\begin{figure}
\begin{center}
  \includegraphics[height=.8\textheight]{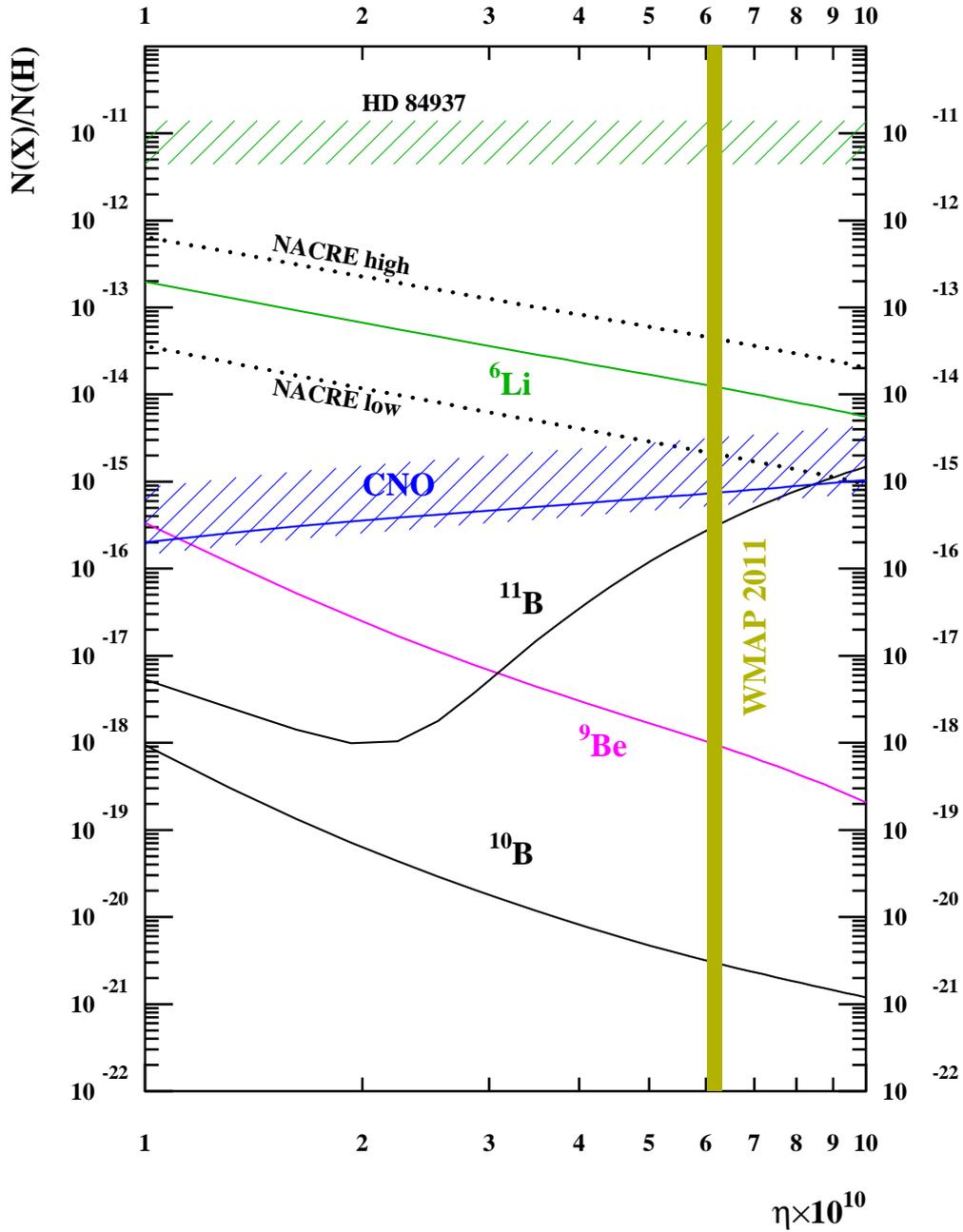}
  \caption{Abundances in number of atoms relative to H for \six, \neu, \dix, \onz\ and CNO isotopes.  For CNO,
  the hatched area represent the present estimated uncertainty. For \six, the dotted lines shows the uncertainty
related to the $^4$He(d,$\gamma)^6$Li reaction before the GSI experiment \cite{Ham10} (solid line )and the   
hatched area shows the most significant \six\ detection in HD 84937 star \cite{Ste10,Ste12}.
  }
\label{f:cnoheli}
\end{center}
\end{figure}

Our sensitivity study, with the extended network, also included \hli\ but no new important reaction
was found besides those previously identified \cite{Coc04,Ser04}, essentially the $^7$Be(d,p)2$\alpha$ reaction.
Indeed, if this reaction rate were higher by a factor of $\sim$100, \sep\ abundance would be brought down to the
observed level \cite{Coc04}. An experiment, performed at Louvain--la--Neuve did not find such an enhancement
in the cross-section  {\em integrated} over the Gamow window  \cite{Ang05}. 
Afterwards, Cyburt \& Pospelov \cite{Cyb12} proposed a resonance enhancement of the cross section that
could have been left undetected by this experiment.
Later, a dedicated experiment at Oak Ridge \cite{OMa11} did not find such a resonance, in the $^7$Be+d channel.
Then, Kirsebom \& Davids \cite{Kir11}
pointed out that the properties of the corresponding $^9$B level had been measured \cite{Sch11}.
When used in the reaction rate and subsequent BBN calculation, the \sep\ depletion was found insignificant \cite{Kir11} ($<$4\%).
The $^7$Be+\tro\ channel was found promising by Chakraborty et al. \cite{Cha11} since the spectroscopy of
the compound nucleus $^{10}$C is deficient in the Gamow window. But as in the $^7$Be+d channel, the required
level properties are at the fringe of standard nuclear physics, as shown by Broggini et al. \cite{Bro12}.
In addition, "missing"  $^{10}$C levels, were not found in a dedicated experiment, recently carried out in Orsay \cite{fairouz}.  
It seems, then that the possibility of a nuclear solution to the lithium problem, has  been ruled out.

\subsection{Results}
\label{s:results}
\bigskip

For  the few reactions that were identified to have an impact on \sbbn, we (re-)evaluated the rates
issued from TALYS, or other sources whose references are listed in the last columns of Table~\ref{t:sensibcno}
to  Table~\ref{t:sensibonz}. 
We were able to collect sufficient experimental data, or theoretical constraints,  to derive new reaction 
rates with associated uncertainties, 
much reduced with respect to our initial three orders of magnitude variation. 
In some cases these new rates differ from the previous ones by large factors but changes compensate
each other (e.g. $^{11}$B(d,n)$^{12}$C and $^{11}$B(d,p)$^{12}$B).
In the meantime, the important $^8$Li($\alpha$,n)$^{11}$B reaction rate, was independently re-evaluated by La Cognata \&  
Del~Zoppo \cite{LaC11}, but it affects CNO production by less than 2\%\footnote{1.4\% indeed, and not a factor of 1.4 (a typo in \cite{Coc12a}).}, 
when compared to our own re-evaluation. 
We hence confirm that the CNO \sbbn\ production is CNO/H $\approx0.7\times10^{-15}$ (number of atoms relative to H). 
Our present analysis does not allow us to precisely quantify
the uncertainty on this result, but from the inspection of Table~\ref{t:sensibcno}, and assuming a factor of ten
uncertainty on the re-evaluated reaction rates, we can estimate the range of CNO/H values to 
$(0.5-3.)\times10^{-15}$. These results are consistent with those of Iocco et al. \cite{Ioc07} but slightly
lower than those of Vonlanthen et al. \cite{Von09}. Detailed CNO, B and Be isotopic abundances, compared with 
those from Iocco et al. \cite{Ioc07} can be found in Table~\ref{t:ycno}, together with H, He and Li isotopic
abundances compared with our previous work. As expected, the extension of the network does not alleviate the
\sep\ discrepancy between calculations and observations. No uncertainty is available yet for the extended 
network result but, for \hli, they are within the reduced network uncertainties \cite{CV10}. The differences in 
the central values are essentially caused by the small evolution of the baryonic density, following the progress
in WMAP data reduction over the years.      


\begin{table}
\begin{center}
\caption{\label{t:ycno} Primordial abundances at (slightly evolving) WMAP baryonic density, with an extended network 
and after reaction rate re-evaluations, compared to previous works.}
\begin{tabular}{ccc}
\hline
   & CV10 \cite{CV10}    & CGXSV \cite{Coc12a} \\
\hline
$Y_p$     & 0.2476$\pm$0.0004       &   0.2476   \\
\deu/H   ($ \times10^{-5})$& $2.68\pm0.15$      &     2.59    \\
\tro/H    ($ \times10^{-5}$) & 1.05$\pm$0.04       &   1.04         \\
\sep/H ($\times10^{-10}$)  &  5.14$\pm$0.50    & 5.24    \\
\six/H ($\times10^{-14}$)& 1.3\cite{Ham10}  & 1.23 \\
\hline
  & Iocco et al. \cite{Ioc07}  &    CGXSV \cite{Coc12a}    \\
\hline
\neu/H ($\times10^{-19}$) &   2.5 &     9.60             \\
\dix/H  ($\times10^{-21}$) & &   3.00  \\
\onz/H  ($\times10^{-16}$) & 3.9    &  3.05      \\
$^{12}$C/H ($\times10^{-16}$)&4.6 & 5.34\\
$^{13}$C/H ($\times10^{-16}$) &0.90& 1.41\\
$^{14}$C/H  ($\times10^{-21}$) &13000. & 1.62\\
$^{14}$N/H ($\times10^{-17}$)&3.7& 6.76 \\
$^{15}$N/H ($\times10^{-20}$)&& 2.25 \\
$^{16}$O/H ($\times10^{-20}$) &2.7& 9.13  \\
CNO/H  ($\times10^{-16}$) &6.00&     7.43 \\
\hline
\end{tabular}
\end{center}
\label{t:abund}
\end{table}

For \six, now that an increased $^2$H($\alpha,\gamma)^6$Li cross section is excluded\cite{Ham10},
the BBN \six\ yield (\six/H$\approx10^{-14}$) at WMAP baryonic density is about two orders 
of magnitude below the reported observations in some halo stars, that nevertheless 
have to be confirmed.

The CNO \sbbn\ production is found to be little sensitive to the baryonic density of the Universe as shown 
on Figure~\ref{f:cnoheli} where the \six, \neu, \dix\ and \onz\ abundances, results of our calculations, are also depicted.
Even when considering our estimated uncertainty, the primordial CNO abundance is too low to have an impact on
Pop III stellar evolution. Those first stars having high masses and consequently short lifetimes are not observed
today. The high C and O abundances observed in extremely low metallicity Pop II stars are expected to have been
synthesized by massive Pop III stars making the determination of primordial CNO abundances from observations   
presently out of reach.

\section{Conclusions}

The baryonic density of the Universe as determined by the analysis of the 
CMB anisotropies is in very good agreement with 
Standard BBN compared to \deu\ primordial abundance
deduced from cosmological cloud observations. However, it 
disagrees with lithium observations in halo stars by a factor that has increased
with the availability of improved nuclear data and astronomical observations. 
Presently, the favored explanation is lithium stellar depletion, 
but the larger needed depletion factor is hardly compatible whit the thin
observed plateau.
It is hence essential to determine precisely the {\em absolute} cross sections
important for \sep\ nucleosymthesis (Table~\ref{t:sensib}). 

Nevertheless, primordial nucleosynthesis remains an invaluable tool for
probing the physics of the early Universe. 
When we look back in time,
it is the ultimate process for which we, {\it a priori}, know all the 
physics involved. Hence, departure from its predictions provide hints
for new physics or astrophysics. 
Gravity could differ from its general relativistic description, for
instance a scalar field, in addition to the tensor field of general 
relativity (GR), appears naturally in superstring theories. 
That would affect the rate of expansion of the universe and hence BBN (see \cite{Coc09} and Ref.~\cite{JPU10} for a review).
Coupled variation of the fundamental couplings is also motivated by 
superstring theories (see Ref.~\cite{JPU11} for a review). 
However, the impact of these variations on the nuclear reaction rates is difficult
to estimate, as in general, nuclear physics uses phenomenological models, whose
parameters are not explicitly linked to fundamental constants. 
The decay of a massive particle during or after BBN could affect
the light element abundances and potentially lower the \sep\ abundance (see e.g.~\cite{Cyb10}).
Negatively charged relic particle, like the supersymmetric partner of the tau lepton, could
form bound states with nuclei, lowering the Coulomb barrier and hence
leading to the catalysis of nuclear reactions (see e.g.~\cite{Pos08,Kus08}). 
Annihilation of dark matter during BBN, e.g. injecting extra neutrons,  could also 
modify the primordial abundances \cite{Jed04,Alb12}.

We have extended our network up to the CNO region and performed
a sensitivity study to identify the few reactions that could affect the A$>$7 isotope yields
and re-evaluated their rates.
The CNO isotope production was found to be in the range CNO/H = $(0.5-3.)\times10^{-15}$, not sufficient to
have an impact on the evolution of the first stars. It is nevertheless a reference value for comparison
with non-\sbbn\ CNO production e.g. in the context of varying constants.
In this particular case, even with a faster triple--alpha reaction rate or a stable $^8$Be, 
the C(NO) production remains $\approx$6 order of magnitude \cite{Coc12a} lower than the \sbbn\ value 
reported here.

Last but not least, we stress here the importance of sensitivity studies in nuclear astrophysics that have been
done, e.g. in the context of novae \cite{Ili02}, X--ray burst \cite{Par08} or massive stars \cite{Ili11}.
Even in the 
simpler context of BBN without the complexity (e.g. mixing) of stellar nucleosynthesis, it would have been
very unlikely to predict the influence of  the $^1$H(n,$\gamma)^2$H reaction on \sep\ nor
of  the $^7$Li(d,n)2$^4$He reaction on CNO.

\ack
I am indebted to all my collaborators on these topics:  Pierre Descouvemont, 
Sylvia Ekstr\"om,  St\'ephane Goriely, Georges Meynet, Keith Olive, Jean-Philippe Uzan, Elisabeth Vangioni and Yi Xu .
This work  was supported in part by the french ANR VACOUL.

\end{document}